\documentclass[12pt]{article}
\usepackage{amsmath}
\usepackage{amsfonts}
\usepackage{amssymb}
\usepackage{amscd}
\usepackage{latexsym,eucal}
\usepackage{graphicx}
\begin{document}
%
%
\def\mkb{\mbox}
\def\beq{\begin{equation}}
\def\eeq{\end{equation}}
\def\beqn{\begin{eqnarray}}
\def\eeqn{\end{eqnarray}}
%
%
\newcommand{\mfootnote}[1]{{}}                          
\newcommand{\mlabel}[1]{\label{#1}}                     
\newcommand{\mcite}[1]{\cite{#1}}                       
\newcommand{\mref}[1]{Eq. (\ref{#1})}
%
\newtheorem{theorem}{Theorem}[section]
\newtheorem{proposition}[theorem]{Proposition}
\newtheorem{definition}[theorem]{Definition}
\newtheorem{lemma}[theorem]{Lemma}
\newtheorem{coro}[theorem]{Corollary}
\newtheorem{prop-def}{Proposition-Definition}[section]
\newtheorem{claim}{Claim}[section]
\newtheorem{remark}[theorem]{Remark}
\newtheorem{example}[theorem]{Example}
\newtheorem{propprop}{Proposed Proposition}[section]
\newtheorem{conjecture}{Conjecture}
\newenvironment{exam}{\begin{example}\rm}{\end{example}}
\newenvironment{rmk}{\begin{remark}\rm}{\end{remark}}
%
%
%
%
\def\ta1{\includegraphics[scale=0.42]{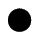}}
\def\tb2{\includegraphics[scale=0.42]{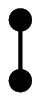}}
\def\tc3{\includegraphics[scale=0.42]{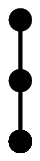}}
\def\td31{\!\!\includegraphics[scale=0.42]{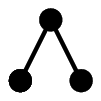}}
\def\te4{\includegraphics[scale=0.42]{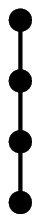}}
\def\tf41{\!\!\includegraphics[scale=0.42]{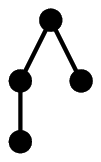}}
\def\tg42{\!\!\includegraphics[scale=0.42]{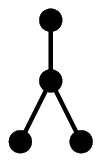}}
\def\th43{\!\!\includegraphics[scale=0.42]{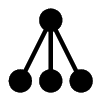}}
\def\ti5{\includegraphics[scale=0.42]{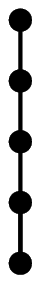}}
\def\tj51{\!\!\includegraphics[scale=0.42]{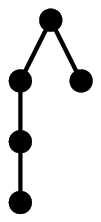}}
\def\tk52{\!\!\includegraphics[scale=0.42]{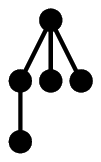}}
\def\tl53{\!\!\includegraphics[scale=0.42]{tree53}}
\def\tm54{\!\!\includegraphics[scale=0.42]{tree54}}
\def\tn55{\!\!\includegraphics[scale=0.42]{tree55}}
\def\tp56{\!\!\includegraphics[scale=0.42]{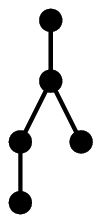}}
\def\tq57{\!\!\includegraphics[scale=0.42]{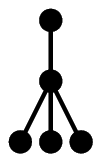}}
\def\tr58{\!\!\includegraphics[scale=0.42]{tree58}}
%
\def\dta{\includegraphics[scale=0.42]{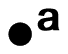}}
\def\dtb{\includegraphics[scale=0.42]{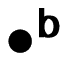}}
\def\dtba{\includegraphics[scale=0.42]{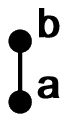}}
\def\dtab{\includegraphics[scale=0.42]{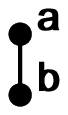}}
\def\graph1{\includegraphics[scale=0.42]{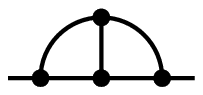}}
%
%
\def\sdta{\includegraphics[scale=0.62]{dectree1}}
\def\sdtb{\includegraphics[scale=0.62]{dectree2}}
\def\sta1{\includegraphics[scale=0.62]{tree1}}
\def\stb2{\includegraphics[scale=0.62]{tree2}}
\def\stc3{\includegraphics[scale=0.62]{tree3}}
\def\std31{\!\!\includegraphics[scale=0.62]{tree31}}
\def\ste4{\includegraphics[scale=0.62]{tree4}}
\def\stf41{\!\!\includegraphics[scale=0.62]{tree41}}
\def\stg42{\!\!\includegraphics[scale=0.62]{tree42}}
\def\sth43{\!\!\includegraphics[scale=0.62]{tree43}}
\def\sti5{\includegraphics[scale=0.62]{tree5}}
\def\stj51{\!\!\includegraphics[scale=0.62]{tree51}}
\def\stk52{\!\!\includegraphics[scale=0.62]{tree52}}
\def\stl53{\!\!\includegraphics[scale=0.62]{tree53}}
\def\stm54{\!\!\includegraphics[scale=0.62]{tree54}}
\def\stn55{\!\!\includegraphics[scale=0.62]{tree55}}
\def\stp56{\!\!\includegraphics[scale=0.62]{tree56}}
\def\stq57{\!\!\includegraphics[scale=0.62]{tree57}}
\def\str58{\!\!\includegraphics[scale=0.62]{tree58}}
%
%
\def\A{\mathcal{A}}
\def\B{{B}^{+}}
\def\BR{\mathcal{B}}
\def\D{\mathcal{D}}
\def\F{\mathcal{F}}
\def\G{\mathcal{G}}
\def\H{\mathcal{H}}
\def\L{\mathcal{L}}
\def\M{\mathcal{M}}
\def\P{\mathcal{P}}
\def\RB{\mathcal{R}}
\def\T{\mathcal{T}}
\def\U{\mathcal{U}}
\def\X{\mathcal{X}}
\def\l{\mathfrak{l}}
\def\g{\mathfrak{g}}
\def\C{\mathbb{C}}
\def\K{\mathbb{K}}
\def\N{\mathbb{N}}
\def\R{\mathbb{R}}
\def\Z{\mathbb{Z}}
\def\to{\rightarrow}
\def\lto{\longrightarrow}
\def\sh{\sqcup \!\! \sqcup}
\def\EXP{exp^{\star}}
\def\EXPR{exp^{\star_R}}
\def\CH{char_{\A}\H_{rt}}
\def\CHR{char_{\A_R}\H_{rt}}
\def\pCH{\partial char_{\A}\H_{rt}}
\def\pCHR{\partial \char_{\A_R}\H_{rt}}
\def\Lie{\mathcal{L}_{\H_{rt}}}
%
%
%
\begin{center}
{\LARGE{Integrable Renormalization II: the general case}}\\[1cm]
\end{center}
\begin{center}
         KURUSCH EBRAHIMI-FARD\footnote{kurusch@ihes.fr}\\
{\small{
         I.H.\'E.S.\\
         Le Bois-Marie, 35, Route de Chartres\\
         F-91440 Bures-sur-Yvette, France\\
         {\small{and}}\\
         Universit\"at Bonn -
         Physikalisches Institut\\
         Nussallee 12,
         D-53115 Bonn,
         Germany}}
\end{center}
\begin{center}
         LI GUO\footnote{liguo@newark.rutgers.edu}\\
{\small{
         Rutgers University\\
         Department of Mathematics and Computer Science\\
         Newark, NJ 07102, USA}}
\end{center}
\begin{center}
         DIRK KREIMER\footnote{kreimer@ihes.fr and dkreimer@bu.edu, Center for Math.Phys., Boston University.}\\
{\small{
         C.N.R.S.-I.H.\'E.S.\\
         Le Bois-Marie, 35, Route de Chartres\\
         F-91440 Bures-sur-Yvette, France}}
\end{center}
\begin{center}
\today
\end{center}
%
%
%
\begin{abstract}

We extend the results we obtained in an earlier work \cite{KLD}.
The cocommutative case of ladders is generalized to a full Hopf
algebra of (decorated) rooted trees. For Hopf algebra characters
with target space of Rota-Baxter type, the Birkhoff decomposition
of renormalization theory is derived by using the Rota-Baxter
double construction, respectively Atkinson's theorem.
We also outline the extension to the Hopf algebra of Feynman
graphs via decorated rooted trees.

\end{abstract}

\section{Introduction}

The perturbative approach to quantum field theory (QFT) has been
spectacularly successful in the past. It is  based on a priori
formal series expansions of Green functions in orders of a
coupling constant, measuring the strength of the corresponding
interaction.
Terms in these series expansions are indexed by Feynman diagrams,
a graphical shorthand for the corresponding  Feynman integrals.
Physically relevant quantum field theories when treated
perturbatively develop short-distance singularities
present in all superficially divergent contributions to the
perturbative expansion. Renormalization theory \cite{Coll} allows
nevertheless for a consistent way to treat these divergent Feynman
integrals in perturbative QFT.
The intricate combinatorial, algebraic and analytic structure of
renormalization theory within QFT is by now known for almost 70
years. Within the physics community the subject reached its final
and satisfying form through the work of Bogoliubov, Parasiuk,
Hepp, and Zimmermann. It was very recently that one of us in
\cite{Kreim1} discovered a unifying scheme in terms of Hopf
algebras and its duals underlying the combinatorial and algebraic
structure. This Hopf algebraic approach to renormalization theory,
as well as the related Lie algebra structures, were exploited in
subsequent work \cite{Kreim2, BK1, BK2, BK3, CK1, CK2, CK3, KM}.

The focus of an earlier work of us \cite{KLD}\footnote{For the
rest of this work we will cite paper \cite{KLD} by (I).} and this
article is on the algebraic Birkhoff decomposition discovered
first in \cite{Kreim2, CK1, CK2}, and the related Lie algebra of
rooted trees, respectively Feynman graphs \cite{CK3, KM}. The
Rota-Baxter algebra structure on the target space of (regularized)
Hopf algebra characters showed to be of crucial importance with
respect to the Birkhoff decomposition
(in \cite{Kreim2} this relation appeared under the name multiplicativity constraint).\\
Using a classical r-matrix ansatz, coming simply from the
Rota-Baxter map, we were able to derive in (I) the formulae for
the factors $\phi_{\pm}$ \cite{CK1} for the decomposition of a
Hopf algebra character $\phi$ in the case of the Hopf subalgebra
of rooted ladder trees. Bogoliubov's $\bar{R}$-map finds its
natural formulation in terms of a character with values in the double of the
above Rota-Baxter algebra. The counterterm $S_R^{\phi}=\phi_{-}$
and the renormalized character $\phi_{+}$ simply lie in the images
of the group homomorphisms $\RB$, $-\tilde{\RB}=\RB-id$,
respectively, of the Bogoliubov character.

In this work we would like to extend these results to the general
case, i.e. the full Hopf algebra of arbitrary rooted trees. 
The main difference lies in the fact that in the rooted ladder tree case we
worked with a cocommutative Hopf algebra, or dually, with the universal 
enveloping algebra of an abelian Lie algebra.
%
%
In the general case the Lie algebra of infinitesimal characters 
is non-abelian, correspondingly the Hopf algebra is non-cocommutative, 
necessitating a more elaborate treatment due to contributions from the 
Baker-Campbell-Hausdorff (BCH) formula.
The modifications coming from these BCH contributions have to be
subtracted order by order in the grading of the Hopf algebra.
Hence we will define in a recursive manner an infinitesimal
character in the Lie algebra which allows for the Birkhoff
decomposition of the Feynman rules regarded as an element of the
character group of the Hopf algebra. The factors of the derived
decomposition give the formulae for the renormalized character
$\phi_{+}$ and the counterterm $S_R^{\phi}=\phi_{-}$ introduced in
\cite{Kreim1,Kreim2,CK1}. Bogoliubov's $\bar{\RB}$-map becomes a
character with values in the double of the target space Rota-Baxter algebra.
It should be underlined that the above ansatz in terms of an
r-matrix solely depends on the algebraic structure of the Lie
algebra of infinitesimal characters, i.e. the dual of the Hopf
algebra of (decorated) rooted trees or Feynman graphs, and on the
Rota-Baxter structure underlying the target space of characters.
When specializing the Rota-Baxter algebra to be the algebra of
Laurent series with pole part of finite order, this approach
naturally reduces to the minimal subtraction scheme in dimensional
regularization, or to the momentum scheme, which are both widely used
in perturbative QFT and thoroughly explored in 
\cite{Kreim1, Kreim2, BK1, BK2}, which extend to non-perturbative
aspects still using the Hopf algebra \cite{BK3}.

The paper is organized as follows. In the following section, we introduce the notion of
Rota-Baxter algebras and recall some related basic algebraic facts, like the relation to
the notion of classical Yang-Baxter type identities, the double construction and Atkinson's
theorem.

After that, we review the notion of a renormalization Hopf
algebra by introducing the universal object \cite{CK1} for such Hopf
algebras, the Hopf algebra of rooted trees. Having this at our
disposal, generalizations to the Hopf algebra of decorated rooted
trees or Feynman graphs are a straightforward generalization which
we will outline later on. Its dual, containing the Lie group of
Hopf algebra characters, and the related Lie algebra of its
generators, i.e. the infinitesimal characters, is introduced
without repeating the  details which are by now standard
\cite{CK1}.

In section four, which contains the main part of this paper, the
notion of a regularized character is introduced as a character
with values in a Rota-Baxter algebra. Note that Bogoliubov-Parasiuk-Hepp-Zimmermann 
(BPHZ) renormalization falls into this class, even though it makes no use of a 
regulator, but of a Taylor operator on the integrand instead, which provides a 
Rota-Baxter map similarly. This is immediate upon recognizing that
disjoint one-particle irreducible graphs allow for
independent Taylor expansions in masses and momenta.
This allows us to lift the Lie algebra of
infinitesimal characters to a Rota-Baxter Lie algebra, giving the
notion of a classical r-matrix on this Lie algebra. We then review
briefly the results of (I), i.e. the Birkhoff factorization in the
cocommutative case.

Motivated by this result for the simple case of rooted ladder trees,
we solve here the factorization problem for the non-cocommutative Hopf
algebra of rooted trees by defining a BCH-modified infinitesimal
character.

Section four closes with some calculations using the notion of
normal coordinates intended to make the construction of the
modified character in terms of the BCH-corrections more explicit,
and a remark on decorated, non-planar rooted trees and Feynman graphs.
%
%
%
\section{Rota-Baxter algebras: from Baxter to Baxter}

The Rota-Baxter (RB) relation first appeared in 1960 in the work
of the American mathematician Glen Baxter \cite{B}. Later it was
explored especially by the mathematicians F. A. Atkinson, G.-C.
Rota and P. Cartier \cite{A, Rota1, C}. In particular, Rota
underlined its importance in various fields of mathematics,
especially within combinatorics \cite{Rota3}. But it was very
recently that after a period of dormancy it showed to be of
considerable interest in several so far somewhat disconnected
areas like Loday type algebras \cite{AguLoday, KEF, PhL1, LG3,
Agu1}, q-shuffle and q-analoga of special functions through the
Jackson integral \cite{Rota2}, differential algebras \cite{LG1,
LG2}, number theory \cite{LG4}, and the Hopf
algebraic approach to renormalization theory in perturbative QFT
\cite{KLD, Kreim2, CK2}. In particular, in collaboration with Connes, 
the connection to Birkhoff decompositions based on Rota--Baxter maps
was introduced in \cite{CK2,CK3}. It is the latter aspect on which we
will focus in this work.

In its Lie algebraic version the RB relation found one of its most
important applications within the theory of integrable systems,
where it was rediscovered in the 1980s under the name of (operator
form of the) classical and modified classical
Yang-Baxter\footnote{Referring to C. N. Yang and the Australian
physicist Rodney Baxter.} equation \cite{STS1, STS2, BD}. There
some of its main features, already mentioned in \cite{A}, and
Atkinson's theorem itself, were analyzed in greater detail.
Especially the double construction introduced in the work of
Semenov-Tian-Shansky \cite{STS1, STS2}, and the related
factorization theorems \cite{STS3} will be of
interest to us.

In the following we will collect a few basic results on Rota-Baxter algebras, some of which we will need
later, and some of which we state just to indicate interesting relations of these algebras to other areas
of mathematics. Of course, the list is by no means complete, and a more exhaustive treatment needs to be
done.

Let us start with the definition of a Rota-Baxter algebra \cite{Rota3, LG1, LG2}.
Suppose $\K$ is a field of characteristic $0$.
A $\K$-algebra neither needs to be associative, nor commutative, nor unital unless stated otherwise.
\begin{definition}
Let $\A$ be a $\K$-algebra with a $\K$-linear map $R: \A \to \A$.
We call $\A$ a Rota-Baxter (RB) $\K$-algebra and $R$ a Rota-Baxter map
(of weight $\theta  \in \K$) if the operator $R$ holds the following
Rota-Baxter relation of weight $\theta \in \K$
\footnote{Some authors denote this relation in the form
$R(x)R(y)=R\big(R(x)y + xR(y)+\lambda xy\big)$. So $\lambda=-\theta$.}:
\beq
    R(x)R(y) + \theta R(xy) = R\big(R(x)y + xR(y)),\ \forall x,y \in \A.
\eeq
\end{definition}
\begin{rmk}
1) For $\theta \ne 0$ a simple scale transformation $R \to \theta^{-1}R$ gives the so called standard form:
\beq
    R(x)R(y) + R(xy) = R\big(R(x)y + xR(y)).    \mlabel{RBR}
\eeq
For the rest of the paper we will always assume the Rota-Baxter map to be of weight $\theta=1$, i.e. to be
in standard form.\\[0.2cm]
2) If $R$ fulfills relation (\ref{RBR}) then $\tilde{R}:=id-R$ fulfills the same Rota-Baxter relation.\\[0.2cm]
3) The images of $R$ and $id-R$ give subalgebras in $\A$.\\[0.2cm]
4) The free associative, commutative, unital RB algebra is given by the mixable shuffle algebra~\cite{LG1} 
which is an extension of Hoffman's quasi-shuffle algebra~\cite{Ho,KLq}. \\[0.2cm]
5) The case $\theta=0$,  $R(x)R(y) = R\big(R(x)y + xR(y))$, naturally translates into the ordinary shuffle
relation, and finds its most prominent example in the integration by parts rule for the Riemann integral.\\[0.2cm]
6) A relation of similar form is given by the associative Nijenhuis identity \cite{CGM}:
\beq
    N(x)N(y) + N^2(xy) = N\big(N(x)y + xN(y)\big).
    \mlabel{ANR}
\eeq
Given a RB algebra with an idempotent RB map $R$, the operator
$N_{\gamma}:=R-\gamma \tilde{R},\;\gamma \in \K$ fulfills relation
(\ref{ANR}). See \cite{LG3, KEF2, PhL2} for recent results with
respect to this relation.
\end{rmk}
\begin{exam}
1) The $q$-analog of the Riemann integral, or Jackson-integral \cite{Rota3, Rota2},
on a well chosen function algebra $\mathcal{F}$ is given by:
\beqn
    J[f](x) &:=& \int_{0}^{x} f(y) d_qy                \nonumber\\
            &:=& (1-q)\:\sum_{n \ge 0} f(xq^n) xq^n.   \label{JI}
\eeqn
It may be written in a more algebraic version, using the operator:
\beq
    P_q[f]:=\sum_{n>0}E_q^{n}[f],         \label{P}
\eeq
where $E_q[f](x):=f(qx),\; f \in \F$.
$P_q$ and $id+P_q=:\hat{P}_q$ are RB operators of weight $-1,\;1$, respectively.
Now let us define a multiplication operator $M_f : \F \to \F,\; f \in \F$,
$M_{f}[g](x):=[f\: g](x)=f(x)g(x)$ which fulfills the associative Nijenhuis relation (\ref{ANR}).
The Jackson integral is given in terms of the above operators as:
\beq
     J[f](x) = (1-q) \hat{P}_qM_{id}\:[f](x),                  \label{qInt}
\eeq
and fulfills the following mixed RB relation
\beq
    J[f]\: J[g] + (1-q)JM_{id}[f \: g] = J \Big[J [f] \: g + f \: J[g] \Big].
\eeq
In a forthcoming work two of us (K.E.-F., L.G.)  will report some
interesting implications of this fact with respect
to some recent results on q-analog of multiple-zeta-values \cite{KLq}.\\[0.2cm]
2) A rich class of Rota-Baxter maps is given by certain projectors.
Within renormalization theory, dimensional regularization together with the minimal subtraction scheme
play an important r{\^o}le. Here the RB map $R_{MS}$ is of weight $\theta=1$ and defined on the algebra
of Laurent series $\C\big[\!\big[\epsilon,\epsilon^{-1}]$ \cite{Kreim2} with finite pole part.
For $\sum_{k=-m}^{\infty}c_k\epsilon^{k} \in \C\big[\!\big[\epsilon,\epsilon^{-1}]$ it gives:
\beq
    R_{MS} \big(\sum_{k = -m}^{\infty}c_k\epsilon^{k}\big):=\sum_{k=-m}^{-1}c_k\epsilon^{k}.
    \mlabel{Rms}
\eeq
\end{exam}
Of equal importance is the projector which keeps the finite part,
closely related to the momentum scheme.

We now introduce the modified Rota-Baxter relation. Its Lie algebraic version
already appeared in \mcite{STS1, BD}.
\begin{definition}
Let $\A$ be a Rota-Baxter algebra, $R$ its Rota-Baxter map.
Define the operator $B: \A \to \A,\;\; B:=id-2R$ to be the modified Rota-Baxter map and
call the corresponding relation fulfilled by $B$:
\beq
    B(x)B(y) = B\big(B(x)y + xB(y)\big) - xy,\ \forall x,y \in \A
    \mlabel{mRBR}
\eeq
the modified Rota-Baxter relation.
\end{definition}
\begin{rmk}
In the following proposition (\ref{LieRB}), we mention the notion of pre-Lie algebras.
Let us state briefly its definition. A (left) pre-Lie $\K$-algebra $\A$ is a $\K$-vector space,
together with a bilinear pre-Lie product $\centerdot: \A \times \A \to \A$, holding the (left) pre-Lie 
relation:
$$
  a \centerdot (b \centerdot c) - (a \centerdot b) \centerdot c 
  = b \centerdot ( a \centerdot c) - (b \centerdot a) \centerdot c ,\;\;\; \forall a,b,c \in \A. 
$$
The commutator $[a,b]:=a \centerdot b - b \centerdot a, \;\; \forall a,b \in \A$ fulfills the Jacobi identity.
\end{rmk}
\begin{proposition} \label{LieRB}
For the Rota-Baxter algebra $\A$ to be either an associative or
pre-Lie $\K$-algebra, the (modified) Rota-Baxter relation
naturally extends to the Lie algebra $\L_{\A}$ with commutator
bracket $[x,y]:=xy-yx,\; \forall x,y \in \A$:
\beqn
    \mlabel{LieRBR}  [R(x),R(y)]\! &\!\!+\!\!&\! R([x,y]) = R\big([R(x),y] + [x,R(y)]\big) \\
    \mlabel{mLieRBR} [B(x),B(y)]\! &\!\!=\!\!&\! B\big([B(x),y] + [x,B(y)]\big) - [x,y].
\eeqn
\end{proposition}
The proof  is a straightforward calculation. The relations
(\ref{LieRBR}) and (\ref{mLieRBR}) are well-known as the (operator
form of the) classical Yang-Baxter and modified Yang-Baxter
equation, respectively.
\begin{rmk}
1) The same is true for the associative Nijenhuis relation (\ref{ANR}). In its Lie algebraic version,
identity (\ref{ANR}) was investigated in \cite{GS, KSM}.\\[0.2cm]
2) Let $\A$ be an associative $\K$-algebra. We regard $\A \otimes \A$ as an $\A$-bimodule,
$x \otimes y \in \A \otimes \A$ and $a(x \otimes y)b = (ax \otimes y)b = ax \otimes yb$.
A solution $r:=\sum_{i} s_{(i)}\otimes t_{(i)} \in \A \otimes \A$ of the extended associative
classical Yang-Baxter relation:
$$
  r_{13}r_{12} -  r_{12}r_{23} + r_{23}r_{13}= \theta r_{13},\;\; \theta\in \K
$$
gives a RB map $\beta: \A \to \A$ of weight $\theta$, defined by
$\beta(x):=\sum_{i} s_{(i)}\:x\:t_{(i)}$. The notation $r_{ij}$
means, for instance, $r_{13}:= \sum_{i} s_{(i)}\otimes 1 \otimes
t_{(i)}$. This example implies many more interesting results with respect to unital
infinitesimal bialgebras, which will be presented elsewhere. The case
$\theta=0$ was already treated in \cite{Agu1}, implying a RB map of weight 0.\\
\end{rmk}

Atkinson gave in \cite{A} a very nice characterization of general RB $\K$-algebras in terms of a so called
subdirect Birkhoff decomposition:
\begin{theorem}
{\rm (Atkinson \mcite{A}):}
For a $\K$-algebra $\A$ with a linear map $R: \A \to \A$ to be a Rota-Baxter $\K$-algebra,
it is necessary and sufficient that $\A$ has a subdirect Birkhoff decomposition.
\mlabel{thm:atk}
\end{theorem}
The proof of this theorem may be found in \cite{A} and will not be given here.
Essentially, the subdirect Birkhoff decomposition in this case means that the Cartesian product
$\D:=(R(\A),-\tilde{R}(\A)) \subset \A\times\A$ is a subalgebra in $\A \times \A$ and that every
element $x \in \A$ has a unique decomposition $x=R(x)+\tilde{R}(x)$.
This should be compared to the results in the Lie algebra case (\ref{LieRBR}) to be found in
\cite{STS1, STS2, STS3, BBT}.

We come now to one of the main facts about RB algebras. In the
following we assume every RB algebra $\A$ to be either an associative algebra 
or a pre-Lie or Lie
algebra. The RB relation then implies furthermore a possibly
infinite hierarchy of the same RB structure in each of the former
cases. We call this the double construction of the RB-hierarchy on
the RB algebra $\A$, given as follows.
\begin{proposition}
Let $\A$ be a Rota-Baxter algebra with (modified) Rota-Baxter map
$R$, set $B=id-2R$. Equipped with the new product:
\beqn
  \mlabel{double}  a \ast_R b &:=& R(a)b + aR(b) - ab             \\
                              &=& -\frac{1}{2}\big(B(a)b+aB(b)\big),
\eeqn
$\A$ is again a Rota-Baxter algebra of the same type, denoted by $\A_R$.
\mlabel{pp:double}
\end{proposition}
The proof of this proposition is immediate by the definition of $\ast_R $.
Following the terminology in \cite{STS1, STS2}, we call this new Rota-Baxter 
algebra $\A_R$ the double of $\A$. It is also in \cite{STS1} where already 
the notion of the double structure for associative algebras equipped with a
modified Rota-Baxter operator was suggested.

\begin{rmk}
1) Let $\A$ be an associative RB algebra. The composition $a \diamond b:= R(a)b - bR(a) + ab$ defines
a pre-Lie structure on $\A$. This aspect becomes more apparent in the context of Loday's
dendriform structures, for which associative RB algebras give a rich class of interesting examples, see
\cite{AguLoday, KEF, PhL1, LG3, Agu1} and references therein.\\[0.2cm]
2) It is obvious that Proposition~\ref{pp:double} implies a
whole, possibly infinite, hierarchy of doubles $\A^{(i)}_R$ (here,
$\ast=\ast_R^{(0)}$ and $\ast_R=\ast_R^{(1)} $):
$$
  \A^{(0)}_R:=\A,\; \A^{(1)}_R:=(\A, \ast_R), \; \cdots, \; \A^{(i)}_R:=(\A, \ast^{(i)}_R),\cdots
$$
$$
  a \ast^{(i)}_R b := \frac{d^i}{dt^i}_{|_{t=0}} e^{-\frac{1}{2}tB}(a)\: e^{-\frac{1}{2}tB}(b),\; a,b \in \A.
$$
Let us call $\A^{(i)}_R$ the i-$th$ double of $\A$, or equivalently the double of $\A^{(i-1)}_R$.
The following diagram serves to visualize the so-called RB-hierarchy:
$$
    \A \xrightarrow{*^{(1)}_R} \A^{(1)}_{R} \xrightarrow{*^{(2)}_R} \A^{(2)}_{R}
                                            \xrightarrow{*^{(3)}_R} \A^{(3)}_{R} \rightarrow \cdots
$$
3) The RB-hierarchy becomes cyclic of period 2 at level $i=3$, for $R$ being an idempotent RB map, 
$R^2=R$, i.e. the $k$-th double product $*^{(k)}_R = *^{(k+2)}_R$.\\[0.2cm]
4) The Rota-Baxter map $R$ becomes an $\K$-algebra homomorphism between $\A^{(i)}_R$ and
$\A^{(i-1)}_R$, $i \in \N$:
\beq
     R(a \ast^{(i)}_R b)=R(a) \ast^{(i-1)}_R R(b). \label{Rhom1}
\eeq
\mlabel{rk:double}
5) For the Rota-Baxter map $\tilde{R}:=id - R$, we have
\beq
  \tilde{R}(a \ast^{(i)}_R b)=-\tilde{R}(a) \ast^{(i-1)}_R \tilde{R}(b). \label{Rhom2}
\eeq
We therefore have the following diagram of $\K$-algebra homomorphisms:
$$
    \A \xleftarrow{R,\tilde{R}} \A^{(1)}_{R} \xleftarrow{R,\tilde{R}} \A^{(2)}_{R}
                                             \xleftarrow{R,\tilde{R}} \A^{(3)}_{R} \leftarrow \cdots
$$
\end{rmk}

We introduce the following composition, using the shuffle product notion formally.
For an associative $\K$-algebra $\A$ and $a,b\in \K$, 
we define \mkb{$a \sh_{\A} b := ab + ba$}. 
For fixed $a_i,b_j\in \A,\, 1\leq i\leq m,\, 1\leq j\leq n$, define recursively
\begin{eqnarray*}
        \lefteqn{(a_1a_2\cdots a_m)\sh_{A} (b_1b_2\cdots b_n)} \\
        &=a_1((a_2\cdots a_m)\sh_{\A} (b_1\cdots b_n))
            +b_1((a_1\cdots a_m)\sh_{\A} (b_2\cdots b_n)), a_i,b_j\in \A.
\end{eqnarray*}
%
\begin{proposition}
Let $\A$ be an associative Rota-Baxter algebra. For $n \in \N, \; x \in \A$ we have
\begin{itemize}
\item[1)]  integer powers of $R(x)$ and $\tilde{R}(x)$ can be written explicitly as:
\beqn
       (-R(x))^n &=&  (-1)^nR \big( x^{\ast_R \:n} \big)                                        \nonumber \\
                 &=& -R \big( x^n + \sum_{k=1}^{n-1} (-R(x))^{n-k} \sh_{\A} \; x^k \big)             \mlabel{Rshuf2} \\
  \tilde{R}(x)^n &=& \tilde{R} \big( (-1)^{(n-1)} x^{\ast_R \:n} \big)                          \nonumber \\
                 &=& \tilde{R} \big( x^n + \sum_{k=1}^{n-1} (-R(x))^{n-k} \sh_{\A} \; x^k \big).     \mlabel{Rshuf2a}
\eeqn

\item[2)]
for $\A$ also being commutative the above formulae simplify to:
\beqn
        (-R(x))^n &=& -R \big( x^n + \sum_{k=1}^{n-1}{n \choose k} (-R(x))^{(n-k)} \: x^k \big),        \mlabel{Rshuf1}\\
   \tilde{R}(x)^n &=& \tilde{R} \big( x^n + \sum_{k=1}^{n-1}{n \choose k} (-R(x))^{(n-k)} \: x^k \big). \mlabel{Rshuf1a}
\eeqn
\end{itemize}
\mlabel{pp:sha}
\end{proposition}
The proof of this proposition follows by induction on $n$.
%

\section{The Hopf algebra of Rooted Trees}

Rooted trees naturally give a convenient way to denote the
hierarchical structure of subdivergences appearing in a Feynman
diagram \cite{Kreim1}, and the structure maps of their Hopf
algebras describe the combinatorics of renormalization of local
interactions, encapsulating Zimmermann's forest formula. For a
renormalizable theory, the hierarchy of subdivergences can always
be resolved into decorated rooted trees (the parenthesized words
of \cite{Kreim1}) upon resolving overlapping divergences using
maximal forests \cite{overl} corresponding to Hepp sectors. This
amounts to a determination of the closed Hochschild one-cocycles
of the Hopf algebra of renormalization for a given quantum field
theory. This is always possible as the rooted trees Hopf algebra
with its one-cocycle $B_+$ is the universal object \cite{CK1} of
graded commutative Hopf algebras. Hence it suffices to study this
universal object, while the details of a specific Hopf algebra of
renormalization of a chosen quantum field theory only provide
additional notational excesses, albeit cumbersome, see
\cite{BK1, BK2, BK3} for applications.

The main ingredient of this universal commutative Hopf algebra of rooted trees is
given by a well-suited non-cocommutative coproduct, defined in
terms of admissible cuts on these rooted trees. The aforementioned
forest formula is then given essentially by the recursively
defined antipode of this Hopf algebra, coming for free from
mathematical structure.
\begin{center}
              \includegraphics[scale=0.6]{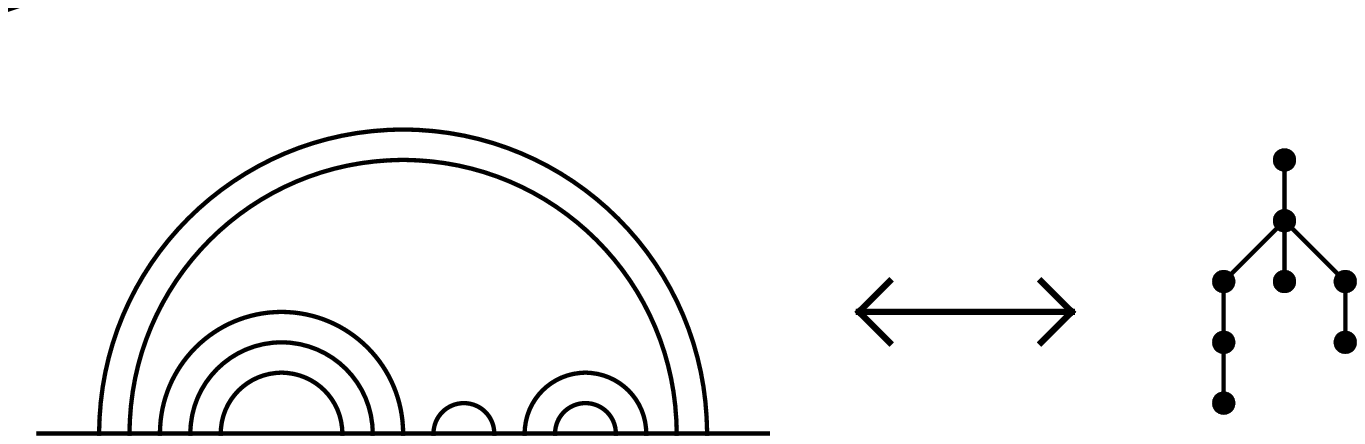}\\
{\small{Fig. 1: A rainbow diagram and corresponding rooted tree of weight 8.}}
\end{center}
Having the Hopf algebra of rooted trees, organizing the algebraic and combinatorial
aspects of renormalization, the description of the analytical structure in terms of
the group of so-called (regularized) Hopf algebra characters takes place within the
dual of this Hopf algebra, being an associative algebra with respect to convolution.

Let us introduce the Hopf algebra of rooted trees \cite{CK1, FGV, M}, which we will denote 
as $\H_{rt}$. The base field $\K$ is assumed once and for all to be of characteristic zero.
By definition a rooted tree $T$ is made out of vertices and nonintersecting
oriented edges, such that all but one vertex have exactly one incoming line.
We denote the set of vertices and edges of a rooted tree by $V(T)$, $E(T)$ respectively.
The root is the only vertex with no incoming line. Each rooted tree is effectively
a representative of an isomorphism class, and the set of all isomorphism classes will
be denoted by $\T_{rt}$.
$$
  \sta1 \;\;\;\; \stb2 \;\;\;\; \stc3 \;\;\;\; \std31 \;\;\;\; \ste4 \;\;\;\; \stf41
  \;\;\;\; \stg42 \;\;\;\; \sth43 \;\;\;\; \sti5 \;\;\;\; \stj51 \;\;\;\; \stk52
  \;\;\;\; \cdots \;\;\;\; \stp56 \;\;\;\; \stq57 \;\;\; \cdots
$$
\begin{definition}
The commutative, unital, associative $\K$-algebra of rooted trees $\A_{rt}$ is the
polynomial algebra, generated by the symbols $T$, each representing an isomorphism
class in $\T_{rt}$. The unit is the empty tree, denoted by $1$, and the product of
rooted trees is denoted by concatenation, i.e. $m_{\A_{rt}}(T,T')=:TT'$.
\end{definition}
We define a grading on the rooted tree algebra $\A_{rt}$ in terms of the number of vertices
of a rooted tree, $\#(T):=|V(T)|$.
This is extended to monomials, i.e. so called forests of rooted trees, by
$\#(T_1 \cdots T_n):=\sum_{i=1}^{n}\#(T_i)$, So that 
$\A_{rt}=\bigoplus_{n \ge 0} \A^{(n)}_{rt}$ becomes a graded, connected, unital, commutative, 
associative $\K$-algebra.\\
Let us introduce now the notion of admissible cuts on a rooted
tree. A cut $c_T$ of a rooted tree is a subset of the set of edges
of $T$, $c_T \subset E(T)$. It becomes an admissible cut, if and
only if along a path from the root to any of the leaves of the
tree $T$, one meets at most one element of $c_T$. By removing the
set $c_T$, $E(T)-c_T$, each admissible cut $c_T$ produces a
monomial of pruned trees, denoted by $P_{c_T}$. The rest, which is
a rooted tree containing the original root, is denoted by
$R_{c_T}$. We exclude the cases, where $c_T=\emptyset$, such that
$R_{c_T}=T,\; P_{c_T}=\emptyset$ and the full cut, such that
$R_{c_T}=\emptyset,\; P_{c_T}=T$. We extend the rooted tree
algebra $\A_{rt}$ to a bialgebra $\H_{rt}$ by defining the counit
$\epsilon: \H_{rt} \to \K$:
\beq
\label{counit}
  \epsilon(T_1 \cdots T_n):= \begin{cases}
                             0 & \;T_1\cdots T_n \ne 1\\
                             1 & \;else.
                             \end{cases}
\eeq
The coproduct $\Delta:\H_{rt} \to \H_{rt} \otimes \H_{rt}$ is defined in terms of
the set of all admissible cuts $C_T$ of a rooted tree $T$:
\beq
    \Delta(T)= T \otimes 1 + 1 \otimes T + \sum_{c_T \in C_T} P_{c_{T}} \otimes R_{c_T}.  \mlabel{coprod1}
\eeq
It is obvious, that this coproduct is non-cocommutative. We extend this by definition to an
algebra morphism.
\begin{definition}
The graded connected Hopf algebra $\H_{rt}:=(\A_{rt}, \Delta,
\epsilon)$ is defined as the algebra $\A_{rt}$ equipped with the
above defined compatible coproduct $\Delta:\H_{rt} \to \H_{rt}
\otimes \H_{rt}$ (\ref{coprod1}), and counit $\epsilon: \H_{rt}
\to \K$ (\ref{counit}).
\end{definition}
\begin{rmk}
1) The coproduct can be written in a recursive way, using the
$\B$ operator, which is a Hochschild 1-cocycle \cite{Kreim2, CK1, FGV}:
\beq
    \Delta(\B(T_{i_1} \cdots T_{i_n}))= T \otimes 1 + \{id \otimes \B\}\Delta(T_{i_1} \cdots T_{i_n}). \label{coprod2}
\eeq
$\B: \H_{rt} \to \H_{rt}$ is a linear operator, mapping a (forest, i.e. monomial of) rooted tree(s) to a rooted
tree, by connecting the root(s) to a new adjoined root:
$$
  \B(1)=\ta1,\;\;\; \B(\ta1)=\tb2,\;\;\; \B(\ta1 \ta1)= \td31,\;\;\; \B(\ta1 \ta1 \ta1)= \th43,
                \;\;\; \B( \tb2 \ta1 \ta1 )=\tk52\;\;\; \cdots
$$
It therefore raises the degree by $1$. Every rooted tree lies in
the image of the $\B$ operator. Its conceptual importance with
respect to fundamental notions of physics was illuminated recently
in \cite{Houches}.\\[0.2cm]
2)
The Hopf algebra $\H_{rt}$ contains a commutative, cocommutative Hopf subalgebra $\H^{l}_{rt}$, generated by the
so called rooted ladder trees, denoted by the symbol $t_n,\; n \in \N $ and recursively defined in terms of the
$\B$ operator, $t_0:=1$, $t_m={\B}^{m}(1)$. The coproduct (\ref{coprod1}) therefore can be written as:
\beq
    \Delta(t_n)=t_n \otimes 1 + 1 \otimes t_n + \sum_{i=1}^{n-1}  t_i \otimes t_{n-i}.
    \mlabel{ladcop}
\eeq
\end{rmk}

The bialgebra $\H_{rt}$ actually is a graded connected Hopf algebra, since due to its grading and connectedness,
it comes naturally equipped with an antipode $S: \H_{rt} \to \H_{rt}$, recursively defined by:
\beq
    S(T):= -T - \sum_{c_T \in C_T} S(P_{c_{T}})R_{c_T}.
    \mlabel{antipode}
\eeq

We come now to the dual $\H^*_{rt}$ of the Hopf algebra of rooted trees, i.e. linear
maps from $\H_{rt}$ into $\K$. It is convenient to denote $f(T)=:\langle f,T \rangle \in \K$,
$f \in \H^*_{rt},\; T\in \H_{rt}$. Equipped with the convolution product:
\beqn
  f \star g (T) &:=& m_{\K}(f \otimes g)\Delta(T),\;\;\; T \in \H_{rt}.                               \label{conv}\\
                &\stackrel{(\ref{coprod1})}{=}& f(T) + g(T) + \sum_{c_T \in C_T} f(P_{c_T})g(R_{c_T}) \nonumber
\eeqn
$$
  \H_{rt} \xrightarrow{\Delta} \H_{rt} \otimes \H_{rt}  \xrightarrow{f \otimes g} \K \otimes \K
          \xrightarrow{m_{\K}} \K
$$
it becomes an associative $\K$-algebra. Its unit is given by the counit $\epsilon$.
\begin{rmk}
Higher powers of the convolution product are defined as follows:
\beq
    f_1 \star f_2 \star \cdots \star f_n := m_{\K}( f_1 \otimes f_2 \otimes \cdots \otimes f_n)\Delta^{(n-1)}
\eeq
$$
  \Delta^{(0)}:=id, \;\;\; \Delta^{(k)}:=(id \otimes \Delta^{(k-1)}) \circ \Delta.
$$
\end{rmk}

$\H^*_{rt}$ contains the set $char_{\K}\H_{rt}$ of Hopf algebra characters, i.e. multiplicative
linear maps with values in the field $\K$.
\begin{definition}
\label{defchar}
A linear map $\phi: \H_{rt} \to \K$ is called a character if
$\phi(T_1T_2)=\phi(T_1)\phi(T_2)$, $T_i \in \H_{rt},\; i=1,2$, i.e. $\phi(1)=1_{\K}$.
We denote the set of characters by $char_{\K}\H_{rt}$.
\end{definition}
\begin{proposition}
\label{groupchar}
The set of characters $char_{\K}\H_{rt}$ forms a group with respect to the convolution product (\ref{conv}).
The inverse of $\phi\in char_{\K}\H_{rt}$ is given in terms of the antipode (\ref{antipode}),
$\phi^{-1}:=\phi \circ S$.
\end{proposition}
\begin{definition}
\label{deri}
A linear map $Z: \H_{rt} \to \K$ is called derivation, or infinitesimal character if
$Z(T_1T_2)=Z(T_1)\epsilon(T_2)+\epsilon(T_1)Z(T_2)$, $T_i \in \H_{rt},\; i=1,2$, i.e. $Z(1)=0$.
The set of infinitesimal characters is denoted by $\partial char_{\K}\H_{rt}$.
\end{definition}
\begin{lemma}
For any $Z \in \partial char_{\K}\H_{rt}$ and $T \in \H_{rt}$ of degree $\#(T)=n<\infty$, we have
for $m>n$, $Z^{\star m}(T)=0$.
\end{lemma}
\begin{rmk}
\label{exp}
The last result implies that the exponential $\EXP(Z)(T):=\sum_{k \ge 0}\frac{Z^{\star k}}{k!}(T)$,
$Z \in \partial char_{\K}\H_{rt}$, is a finite sum, ending at $k=\#(T)$.
\end{rmk}

The above facts culminate into the following important results \cite{CK1, M}.
Given the explicit base of rooted trees generating $\H_{rt}$, the set of derivations
$\partial char_{\K}\H_{rt}$ is generated by the dually defined infinitesimal characters, indexed
by rooted trees:
\beq
  Z_{T}(T')=\langle Z_T,T'\rangle := \delta_{T,T'}. \label{Z}
\eeq
\begin{proposition}
The set $\partial char_{\K}\H_{rt}$ defines a Lie algebra, denoted by $\Lie$, and equipped with
the commutator:
\beqn
    [Z_{T'},Z_{T''}] &:=& Z_{T'} \star Z_{T''} - Z_{T''} \star Z_{T'}                  \mlabel{Liebra}\\
                     &=&  \sum_{T \in \T_{rt}} \big( n(T',T'';T)-n(T'',T';T) \big)Z_T, \nonumber
\eeqn
where the $n(T',T'';T) \in \N$ denote so called section coefficients, which count the number
of single simple cuts, $|c_T|=1$, such that $P_{c_{T}}=T'$ and $R_{c_T}=T''$.\\
The exponential map $\EXP: \Lie \to char_{\K}\H_{rt}$ defined in remark (\ref{exp}) is a bijection.
\end{proposition}
Generated by the infinitesimal characters $Z_T$ (\ref{Z}), the Lie algebra $\Lie$ carries naturally
a grading in terms of the grading of the rooted trees in $\H_{rt}$, $deg(Z_T):=\#(T)$, and
$\Lie=\bigoplus_{n>0} \Lie^{(n)}$. The commutator (\ref{Liebra}) implies then:
\beq
    \big[\Lie^{(n)},\Lie^{(m)} \big] \subset \Lie^{(m+n)}. \label{liegrad}
\eeq
Let us calculate a few commutators, to get a better feeling for the structure of $\Lie$:
\beq
 [Z_{\ta1},Z_{\tb2}]  = Z_{\tc3} + 2Z_{\td31} -  Z_{\tc3}  = 2Z_{\td31}  \label{cherry}
\eeq
$$
      [Z_{\ta1},Z_{\tc3}]  = Z_{\te4} + Z_{\tf41} + 2Z_{\tg42} - Z_{\te4} =  Z_{\tf41} + 2Z_{\tg42}
$$
$$
      [Z_{\td31},Z_{\ta1}] =  \frac{1}{2}[[Z_{\ta1},Z_{\tb2}],Z_{\ta1}] = Z_{\tg42} - 3Z_{\th43} - Z_{\tf41}.
$$

This Lie algebra received more attention recently \cite{KM, CK4, Mex1}, but needs further 
structural analysis, since it captures in an essential way the whole of renormalization 
and the structure of the equations of motion \cite{BK3} in perturbative QFT. 
This remark is underlined by the results presented in the next section.

\section{Classical r-Matrix and Birkhoff decomposition}

For a renormalizable theory, the process of renormalization removes the short-distance
singularities order by order in the coupling constant. For this to
work one has to choose a  renormalization scheme which determines
the remaining finite part. This choice is of analytic nature but
also contains an important algebraic combinatorial aspect, which
lies at the heart of the Birkhoff factorization, found in
\cite{Kreim2,CK2}. It is the goal of this section to clarify how
this algebraic step implies the Birkhoff decomposition in a
completely algebraic manner. We derive the corresponding theorem
for graded connected Hopf algebras quite generically. The main
ingredient is a generalized notion of regularization in terms of a
Rota-Baxter structure, which is supposed to underlie the target
space of the characters of $\H_{rt}$.

Following the Hopf algebraic approach to renormalization in
perturbative QFT, we henceforth introduce the notion of
regularized (infinitesimal) characters, maps from $\H_{rt}$ into a
commutative, associative, unital Rota-Baxter algebra $\A$. The
choice of the Rota--Baxter map is determined by the choice of the
renormalization scheme, which can be a BPHZ scheme (Taylor
subtractions of the integrand), the before-mentioned minimal
subtraction and momentum schemes, and others, which all provide
Rota--Baxter maps. Here is not the space to give a complete census
of renormalization schemes in use in physics, but we simply assume
Feynman rules and a Rota--Baxter map being given.\\[-0.2cm]

Let us mention that sometimes we write R-matrix, instead of the standard notation r-matrix, 
to underline its operator form, and origin in the Rota-Baxter relation.\\[-0.2cm]

We therefore generalize $\H^*_{rt}$ to $L(\H_{rt},\A)$, consisting of $\K$-linear maps from
$\H_{rt}$ into the Rota-Baxter algebra $\A$, i.e.
$\langle \phi,T\rangle \in \A,\; \phi \in L(\H_{rt},\A),\; T \in \H_{rt}$.
Due to the double structure on the Rota-Baxter algebra (\ref{double}) we naturally get $L(\H_{rt},\A_R)$.
We then lift the Rota-Baxter map $R: \A \to \A$ to $L(\H_{rt},\A)$, which is possible since it is linear.

\begin{proposition}
Define the linear map $\RB: L(\H_{rt},\A) \to L(\H_{rt},\A)$ by
$f \mapsto \RB(f):=R \circ f: \H_{rt} \to R(\A)$.
Then $L(\H_{rt},\A)$ becomes an associative, unital Rota-Baxter algebra.
The Lie algebra of infinitesimal characters $\Lie \subset L(\H_{rt},\A)$ with bracket
(\ref{Liebra}) becomes a Lie Rota-Baxter algebra, i.e. for $Z',Z'' \in \partial char_{\A}\H_{rt}$,
we have the notion of a classical R-matrix respectively classical Yang-Baxter relation:
\beq
    [\RB(Z'),\RB(Z'')]=\RB\big([Z',\RB(Z'')]\big) + \RB\big([\RB(Z'),Z'']\big) - \RB \big([Z',Z'']\big).
    \mlabel{Lie}
\eeq
\end{proposition}
Notice that we replaced $\K$ by $\A$ for the target space of the regularized infinitesimal characters.
The proof of this proposition was given in (I).

Using the double construction and Atkinson's theorem of Section 2 we have the following
\begin{lemma}
The Rota-Baxter algebra $L(\H_{rt},\A)$ equipped with the convolution product:
\beq
     f \star_{\RB} g  =  f \star \RB(g) + \RB(f) \star g - f \star g
\eeq
gives a Rota-Baxter algebra structure on the set of linear
functionals with values in the double $\A_R$ of $\A$, denoted by
$L(\H_{rt},\A_R)$. An analog for $\Lie$ exists, denoted by
${\Lie}_{\RB}$, equipped with the $\RB$-bracket: 
\beqn
    [Z',Z'']_{\RB} &=& [Z',\RB(Z'')] + [\RB(Z'),Z''] - [Z',Z'']          \nonumber\\
                   &=& \frac{-1}{2}\;\big( [Z',\BR(Z'')] + [\BR(Z'),Z''] \big). \nonumber
    \mlabel{Rbracket}
\eeqn
The $\RB$ map becomes a (Lie) algebra morphism (${\Lie}_{\RB} \to \Lie$) $L(\H_{rt},\A_R) \to L(\H_{rt},\A)$.
\end{lemma}
\begin{rmk}
1)
The above is also true for $\tilde{R}:=id-R$, respectively $\tilde{\RB}:=id-\tilde{\RB}$ (see remark~\ref{rk:double}).\\[0.2cm]
2)
We will denote the Lie subalgebras  $\RB(\L_{\H_{rt}})$ by $\L^{-}_{\H_{rt}}$ and  $\tilde{\RB}(\L_{\H_{rt}})$
by $\L^{+}_{\H_{rt}}$.
\end{rmk}
We now apply Atkinson's theorem to the Lie algebra $\L_{\H_{rt}}$ of infinitesimal characters, the
generators of the group of Hopf algebra characters $char_{\A}\H_{rt}$.

\begin{lemma}
Every infinitesimal character $Z \in \L_{\H_{rt}}$ has a unique subdirect Birkhoff decomposition
$Z=\RB(Z) + \tilde{\RB}(Z)$.
\end{lemma}
\begin{rmk}
1) In the case of an idempotent Rota-Baxter map $R$ we have a direct decomposition $\A=\A_- + \A_+$
respectively $\L_{\H_{rt}}=\L^-_{\H_{rt}} + \L^+_{\H_{rt}}$.\\
2) Let $Z \in \L_{\H_{rt}}$ be the infinitesimal character
generating the character $\phi=exp^{\star}(Z) \in
char_{\A}\H_{rt}$. Using the result in Proposition~\ref{pp:sha},
we then see that for elements in $ker(\epsilon)$, the augmentation
ideal, we have
$$
exp^{\star}(-\RB(Z))=\RB \big( exp^{\star_{\RB}}(-Z) \big),\;\;\;
exp^{\star}(\tilde{\RB}(Z))=-\tilde{\RB} \big( exp^{\star_{\RB}}(-Z) \big).
$$
\end{rmk}

\subsection{Review of the Ladder case}
For the Hopf subalgebra of rooted ladder trees, introduced in the last section,
we found in (I) the following simple factorization for a regularized character
$char_{\A}\H^{l}_{rt} \ni \phi=\EXP(Z),\; Z \in \partial char_{\A}\H^{l}_{rt}$
due to the abelianess of $\L_{\H^{l}_{rt}}$, and induced by Atkinson's theorem, i.e.
the lifted Rota-Baxter map $\RB$:
\beqn
     \phi &=& \EXP(Z)                                                       \\
          &=& \EXP\big(\RB(Z)+\tilde{\RB}(Z)\big)                           \\
          &=& \phi^{-1}_{-} \star \phi_{+}                                  \label{BirkLadder}
\eeqn
where:
\beq
     \phi_{-} = exp^{\star}(-\RB(Z)) \;\;\;\;   \phi_{+} = exp^{\star}(\tilde{\RB}(Z)),      \label{expCKlad}
\eeq
and such that we arrive at the following formulae \cite{CK1} for $\phi_{\pm}$, using Proposition~\ref{pp:sha}:
\begin{proposition}
In the rooted ladder tree case, $\H_{rt}^{l}$, we find for the factors (\ref{expCKlad})
in the Birkhoff decomposition (\ref{BirkLadder}) the following explicit formulae:
\beqn
    \phi_{-}(t_n)  &=& \RB\big(exp^{\star_{\RB}}(-Z)\big)(t_n)                                              \mlabel{A}   \\
                   &=& -R \big\{ \phi(t_n) + \sum_{k=1}^{n-1} \phi_{-}(t_k)\phi(t_{n-k}) \big\}           \mlabel{phi-}\\
    \phi_{+}(t_n)  &=& -\tilde{\RB}\big(exp^{\star_{\RB}}(-Z)\big)(t_n)                                     \mlabel{B}   \\
                   &=& \tilde{R} \big\{ \phi(t_n) + \sum_{k=1}^{n-1} \phi_{-}(t_k)\phi(t_{n-k}) \big\}.   \mlabel{phi+}
\eeqn
\end{proposition}

We emphasize that the map:
\beqn
     b[\phi](t_n) &:=& exp^{\star_{\RB}}(-Z)(t_n)                               \label{bogomap1}\\
                  & =& -\phi(t_n) - \sum_{k=1}^{n-1} \phi_{-}(t_k)\phi(t_{n-k})
\eeqn
which we will call Bogoliubov character, is a Hopf algebra
character $\H^{l}_{rt} \to \A_{R}$, i.e.~into the double
$\A_{R}$ of $\A$. This gives a natural algebraic expression for
Bogoliubov's $\bar{R}$-map. In the next section, where we treat
the general case, we will formally introduce $b[\phi]$. The
Rota-Baxter maps $\RB$ and $-\tilde{\RB}$, i.e. the Lie algebra
homomorphisms, become group homomorphisms 
$char_{\A_{R}}\H_{rt} \xrightarrow{\RB,-\tilde{\RB}}
char_{\A}\H_{rt}$. We will generalize this to arbitrary rooted
trees in the following section, using equation (\ref{Rshuf2a}).
%
%

\subsection{The General Case}

As stated above, in (I) we introduced a classical R-matrix coming
from the Rota-Baxter structure underlying the target space of
regularized characters. We saw that in the case of the Hopf
subalgebra of rooted ladder trees, the abelianess of the related
Lie algebra implies a somehow simple Birkhoff factorization
(\ref{BirkLadder}, \ref{expCKlad}) respectively the formulae
 for the factors $\phi_{\pm}$ (\ref{phi-}, \ref{phi+}).
The general, i.e. non-cocommutative case can be solved due to the graded, connectedness
of the Hopf algebra of rooted trees.

Suppose we start with an infinitesimal character $Z \in \pCH$
generating the regularized Hopf algebra character $\phi=\EXP(Z)
\in \CH$. The above mentioned properties of the Hopf algebra allow
for a recursive definition of an infinitesimal character
$\chi=\chi(Z) \in \pCH$, defined in terms of $Z$, using the lower
central series of Lie algebra commutators. 

Setting 
\beq
     \chi(Z)=Z+\sum_{k=1}^\infty \chi^{(k)}_Z    \label{chi}
\eeq 
we proceed in the following manner. We first introduce the related series 
\beq
    \chi(u;Z) = Z+\sum_{k=1}^\infty u^k\chi^{(k)}_Z, \label{chiu}
\eeq 
where we assume that $0<u<1$ is a real parameter. We also set $\chi^{(0)}_Z=Z$.
We next introduce the Baker--Campbell--Hausdorff (BCH) series  
\beq
    BCH(A,B):= \frac{1}{2}[A,B] + \frac{1}{12}\bigg( \big[A,[A,B]\big] -  \big[B,[A,B]\big] \bigg) + \cdots. \label{ch}
\eeq
See \cite{Varad} for more details on the BCH formula.
Let us write equation (\ref{ch}) in the form 
\beq 
    BCH(A,B)=\sum_{k=1}^\infty c_k K^{(k)}(A,B),
\eeq 
such that the $K^{(k)}$ are the appropriate nested- or multicommutators of depth $k \in \N$, 
i.e. $K^{(1)} = [A,B]$, $K^{(2)} = [A,[A,B]]-[B,[A,B]]$ and so on. Then, the $\chi^{(k)}_Z$
are defined as the solution of the fix point equation 
\beq
    \chi(u;Z) = Z - \sum_{k=1}^{\infty}c_k u^k K^{(k)}\left(\RB(\chi(Z)),\tilde{\RB}(\chi(Z))\right). \label{BCH}
\eeq
Note that $\chi(u;Z)(T)$ is a polynomial in $u$ of degree $m-2$
for any finite tree $T$ of degree $\#(T) = m$ say, i.e.~with $m$ vertices and
therefore is well-defined at $u=1$. We hence set $\chi(Z)\equiv
\chi(1;Z)$. Furthermore, $\chi(Z)$ is an infinitesimal character
as it is a finite linear combination of infinitesimal characters,
and thus the above definition on trees implies its action on
forests as a derivation in the sense of definition  (\ref{deri}).

It is immediate that $\chi^{(k)}_{Z}$ vanishes for all $k \geq 1$
when applied to cocommutative Hopf algebra elements.

Let us work out the cases $k=1,2$ as examples:
\beq
    \chi^{(1)}= - \frac{1}{2}[\RB(Z),\tilde{\RB}(Z)]= - \frac{1}{2}[\RB(Z),Z] \label{1stcorrect}
\eeq
and for $k=2$ we have:
\beqn
    \chi^{(2)} &=& -\frac{1}{2}[\RB(\chi^{(1)}),\tilde{\RB}(Z)] - \frac{1}{2}[\RB(Z),\tilde{\RB}(\chi^{(1)})]         \nonumber\\
         & & \hspace{2cm} - \frac{1}{12}\big( \big[\RB(Z),[\RB(Z),Z]\big] -  \big[\tilde{\RB}(Z),[\RB(Z),Z]\big] \big) \nonumber\\
         &=& +\frac{1}{4}[\RB([\RB(Z),Z]),\tilde{\RB}(Z)] + \frac{1}{4}[\RB(Z),\tilde{\RB}([\RB(Z),Z])]               \nonumber\\
         & & \hspace{2cm}  -\frac{1}{12}\big( \big[\RB(Z),[\RB(Z),Z]\big] -  \big[\tilde{\RB}(Z),[\RB(Z),Z]\big] \big)  \nonumber\\
         &=& \frac{1}{4}\big[\RB([\RB(Z),Z]),Z\big] + \frac{1}{12}\big( \big[\RB(Z),[\RB(Z),Z]\big] - \big[[\RB(Z),Z],Z\big] \big).
              \label{2ndcorrect}
\eeqn 
where $\tilde{\RB}$ has completely vanished.
This nontrivial fact comes partly from $\tilde{\RB}={\rm id}-\RB$, and in a moment we show that the
$\chi^{(k)}_{Z}$ solve the simpler recursion: 
\beq
    \chi(u;Z)=Z + \sum_{k=1}^{\infty}c_k u^k K^{(k)}\left(-\RB(\chi(Z)),Z)\right)\label{BCH1}.
\eeq 
Indeed, from the relation (\ref{BCH}) and the recursive definition of the $\chi^{(k)}_Z$  we have 
the following factorization for group like elements in $\H^*_{rt}$:
\begin{proposition}
\label{CKfactor}
Using the infinitesimal character $\chi \in \pCH$ defined in (\ref{BCH}), we
have the following decomposition of a character $\phi \in \CH$
given in terms of its generating infinitesimal character $Z \in \pCH$:
\beq
     \EXP(Z) = \EXP\big(\RB(\chi(Z))\big) \star \EXP\big(\tilde{\RB}(\chi(Z))\big). \label{FACT}
\eeq
\end{proposition}
This then implies the simpler recursion (\ref{BCH1}), in which the vanishing of $\tilde{\RB}$ is apparent.
\begin{rmk}
\label{BCHformula}
The above formal derivation of the factorization of $\H_{rt}$ characters using the BCH formula in
(\ref{BCH}) and (\ref{BCH1}) to define the infinitesimal character $\chi(Z)$ (\ref{chi}) may be 
summarized in a more suggestive manner by the following two recursive formulae:
\beqn
    \chi(Z) &=& Z - BCH\big(\RB(\chi(Z)),\tilde{\RB}(\chi(Z))\big) \nonumber\\
            &=& Z + BCH\big(- \RB(\chi(Z)),Z\big).                \nonumber  
\eeqn
\end{rmk}

Let us define the factors, i.e. characters $\phi_{\pm} \in \CH$:
\beq
    \phi^{-1}_{-}:= \EXP\big(\RB(\chi(Z))\big), \;\;\;\; \phi_{+}:=\EXP\big(\tilde{\RB}(\chi(Z))\big)
\eeq
and introduce the Bogoliubov character now in general via the following definition.
\begin{definition}
\label{bogo} 
Let $\EXP(Z)=\phi \in \CH$. We define the following character $b[\phi] \in \CHR$ 
with values in the double of $\A$, and call it Bogoliubov character:
\beq
    b[\phi]:=\EXPR(-\chi(Z)).
\eeq
\end{definition}

Remembering the crucial property of exponentiated Rota-Baxter maps, coming from (\ref{double}),
respectively (\ref{Rhom1}, \ref{Rhom2}):
\beqn
     \EXP\big(-\RB(Z)\big) &=& \RB\big(\EXPR(-Z)\big)                     \label{RB1}\\
     \EXP\big(\tilde{\RB}(Z)\big) &=&-\tilde{\RB}\big(\EXPR(-Z)\big),     \label{RB2}
\eeqn
for elements in the augmentation ideal $ker(\epsilon)$, we have:
\beq
    \phi_{-}:=\RB(b[\phi]), \;\;\;\; \phi_{+}:= -\tilde{\RB}(b[\phi]).
\eeq
We use the factorization (\ref{FACT}) in proposition (\ref{CKfactor}) to derive an explicit formula 
for the characters $b[\phi]$ respectively $\phi_{\pm}$.
Let $T \in ker(\epsilon)$, using the coproduct (\ref{coprod1}), we get:
\beqn
     -\tilde{\RB}(b[\phi])(T) &=& \EXP(-\RB(\chi(Z))) \star \EXP(Z)(T)                                         \label{BCH2}\\
                              &=& \EXP(Z)(T) + \RB(b[\phi])(T)                                                 \nonumber\\
                              & & \hspace{1cm} + \sum_{n \ge 0} \frac{1}{n!}\sum_{j=1}^{n-1} {n \choose j}
                                                            \RB(-\chi(Z))^{\star(n-j)} \star Z^{\star j}(T),   \label{a}
\eeqn
where (\ref{BCH2}) again implies the simpler recursion equation (\ref{BCH1}).
\begin{rmk}
1) All expressions are well-defined since they reduce to finite sums for an element 
$T \in ker(\epsilon)$ of finite order $\#(T)=m < \infty$.\\[0.2cm]
2) In the last expression in equation (\ref{a}), the primitive part in $\Delta(T)$ is mapped to zero, since only
strictly positive powers of infinitesimal characters appear.
\end{rmk}

Continuing the above calculation, we get the following:
\beqn
     -\tilde{\RB}(b[\phi])(T) - \RB(b[\phi])(T) &=& -b[\phi](T)                                                  \nonumber \\
                                                &=&  \EXP(Z)(T)                                                  \nonumber \\
                                                & & \hspace{0.6cm} +  \sum_{n \ge 0} \frac{1}{n!} \sum_{j=1}^{n-1} {n \choose j}
                                                                 \RB(-\chi(Z))^{\star(n-j)} \star Z^{\star j}(T),     \nonumber
\eeqn
and therefore we find the well know formula:
\beqn
   \RB\big( \EXPR(-\chi(Z)) \big)(T) &=& -R\big( \EXP(Z)(T) + \nonumber\\
                                     & & \hspace{0.6cm} \sum_{n \ge 0} \frac{1}{n!}\sum_{j=1}^{n-1} {n \choose j}
                                                                       \RB(-\chi(Z))^{\star(n-j)} \star Z^{\star j}(T)\big). \nonumber
\eeqn
Finally, we rederive the results of \cite{Kreim1, Kreim2, CK1} which
gave the counterterm and the renormalized contribution as the
image of the Bogoliubov character under the group homomorphisms $\RB$ and $-\tilde{\RB}$,
now derived from the double construction for any algebraic Birkhoff decomposition based 
on a suitable $\RB$, i.e. Rota-Baxter type map:
\begin{theorem}
\label{CKformula}
For $T \in \H_{rt},\; \#(T)=m$ we have the following formulae for the factors in (\ref{FACT}):
\beqn
%
%
         \RB(b[\phi])(T) &=& \phi_{-}(T)                                                                    \nonumber \\
                         &=& -R\big( \phi(T) + \sum_{n \ge 0}^{m} \frac{1}{n!} \sum_{j=1}^{n-1} {n \choose j}
                                                     \RB(-\chi(Z))^{\star(n-j)} \star Z^{\star j}(T)\big).  \nonumber \\
                         &=& -R\big( \phi(T) + \sum_{c_T \in C_T} \phi_{-}(P_{c_T})\phi(R_{c_T})\big).      \nonumber
\eeqn
and
\beqn
%
%
\tilde{\RB}\big(b[\phi]\big)(T) &=& \phi_{+}(T)                                                           \nonumber \\
                                &=& \tilde{R}\big( \phi(T)
                                      + \sum_{n \ge 0}^{m} \frac{1}{n!}\sum_{j=1}^{n-1} {n \choose j}
                                                  \RB(-\chi(Z))^{\star(n-j)} \star Z^{\star j}(T)\big). \nonumber  \\
                                &=& \tilde{R}\big( \phi(T)
                                      + \sum_{c_T \in C_T} \phi_{-}(P_{c_T})\phi(R_{c_T})\big).           \nonumber
\eeqn
\end{theorem}
This should be compared to the general equation (\ref{Rshuf2}) including the shuffle:
\beqn
     -\EXPR(-\chi(Z))(T) &=& \EXP(Z)(T) + \sum_{n \ge 0}\frac{1}{n!}
                                    \sum_{j=1}^{n-1} {n \choose j} \RB(-\chi(Z))^{\star(n-j)} \star Z^{\star j}(T).     \nonumber \\
                         &=& \EXP(\chi(Z))(T) +                                                                         \nonumber \\
                         & & \hspace{0.6cm} \sum_{n \ge 0}\frac{1}{n!}
                                      \sum_{j=1}^{n-1} \RB(-\chi(Z))^{\star(n-j)} \sh_{\star} \chi(Z)^{\star j}(T).  \label{shufBCH}
\eeqn
It allows us to define the infinitesimal character $\chi=\chi(Z)$ to order $k>0$ in another way recursively
by using the $\chi^{(j)}_Z,\;\; j<k$. We therefore get to order $k$:
\beqn
     \chi^{(k)}_Z &=& -\sum_{j=1}^{k-1}\chi^{(j)}_Z - \sum_{l=1}^{k+2}\frac{1}{l!}\chi(Z)^{\star\, l}  +
                       \sum_{l=1}^{k+2}\frac{1}{l!}Z^{\star\, l}                   \nonumber\\
                & & \;\;\;\;\; - \sum_{n \ge 0}^{k+2}\frac{1}{n!}
                                 \sum_{j=1}^{n-1} \RB(-\chi(Z))^{\star(n-j)} \sh_{\star} \chi(Z)^{\star j}            \nonumber\\
                & & \;\;\;\;\;\;  + \sum_{n \ge 0}^{k+2}\frac{1}{n!}
                                        \sum_{j=1}^{n-1} {n \choose j} \RB(-\chi(Z))^{\star(n-j)} \star Z^{\star j}.
\eeqn

After these formal arguments based on the general results for Rota-Baxter operators and the structure
of the rooted tree Hopf algebra, we end this section and the paper with a remark on calculational aspects. 

When treating the rooted ladder case, we mentioned at the end of (I) the use of normal
coordinates introduced by Chryssomalakos et al. in \cite{Mex1}. 
Given a regularized character $\phi \in \CH$, this provided a very easy way to define the 
coefficients for its generator $Z \in \pCH$:
\beq
    Z:=\sum_{n > 0} \alpha^{(n)}Z_{t_n},\;\; \alpha^{(n)} \in \A \label{ladderZsum}
\eeq
such that $\EXP(Z)(t_n)=\phi(t_n) \in \A$.
\begin{rmk}
1) Here and also later, we omit for notational reasons the tensor sign between
the $\alpha^{(n)}$ and $Z_{t_n}$, i.e. $\alpha^{(n)}Z_{t_n} \in A \otimes \pCH, \; n>0$.\\[0.2cm]
2) The $\alpha^{(n)}$ were given in terms of Schur polynomials, i.e.
$\alpha^{(n)} := \phi(P(t_1,\cdots,t_n))$. And $\phi_{-}$ was just $\EXP(-\RB(Z))$.
\end{rmk}

In the general case, i.e. for arbitrary rooted trees, the simple Schur
polynomials get replaced by the following set of polynomial
equations. Details may be found in \cite{Mex1}. Introducing the
symbols $x^T$, indexed by rooted trees, and defining the new
coordinates, which are characterized by $\EXP(Z)(x^T)=\phi(x^T), \; \phi \in \CH$, where:
\beq
    Z=\sum_{T \in \T_{rt}} \alpha^{x^T}Z_T \in \pCH, \label{Zsum}
\eeq
we arrive at the following infinite set of coupled equations, expressing the coordinates
$T$ in terms of the new $x^T$:
\beq
    T= \sum_{n \ge 0} \frac{1}{n!} m_{\H_{rt}}( \P^{\otimes n+1}) \Delta'^{(n)}(x^T),\;\;\;\; T \in \T_{rt}.
\eeq
Here, the map $\P$ denotes the projector into the augmentation ideal:
\beq
 \P(x^{T_1} \cdots x^{T_n}):= \begin{cases}
                                    \;\;\;\;0, & \hspace{-1cm}x^{T_1} \cdots x^{T_n} = 1\\
                             x^{T_1} \cdots x^{T_n}, & \; else.
                             \end{cases}
\eeq
$\Delta'$ denotes the coproduct reduced to single simple cuts $|c_T|=1$, and
$\Delta'^{(n)}:=(id \otimes \Delta'^{(n-1)})\circ \Delta'$, such that 
$\Delta'^{(0)}:=id, \; \Delta'^{(1)}=\Delta'$. One should compare this
operation with the formal linear map $\EXP(\hat{Z})(T)$ on $\H_{rt}$, where
$\hat{Z}:=\sum_{T \in \T_{rt}} T\:Z_T$, $T\:Z_T(T')=T \delta_{T,T'}$, and
$T_1\:Z_{T_1} \star T_2\:Z_{T_2}:= T_1T_2\:Z_{T_1}\star Z_{T_2}$.\\ 
The first four equations for the rooted trees $\ta1,\; \tb2,\; \tc3,\; \td31$ are:
\beqn
    \ta1  &=& x^{\ta1}                                                           \nonumber \\
    \tb2  &=& x^{\tb2} + \frac{1}{2}x^{\ta1}x^{\ta1}                             \nonumber \\
    \tc3  &=& x^{\tc3} + x^{\ta1}x^{\tb2} + \frac{1}{6}x^{\ta1}x^{\ta1}x^{\ta1},\;\;\;\;\;
    \td31  = x^{\td31} + x^{\ta1}x^{\tb2} + \frac{1}{3}x^{\ta1}x^{\ta1}x^{\ta1}  \nonumber
\eeqn
The final step, done in \cite{Mex1}, is to invert the above equations, giving the $x^T$'s
in terms of the original rooted trees. In the ladder case we just get the Schur polynomials.
In general, we have for example:
\beq
    x^{\tc3}  = \tc3 - \ta1 \tb2 + \frac{1}{3}\ta1 \ta1 \ta1,\;\;\;\;\;
    x^{\td31} = \td31 - \ta1 \tb2 + \frac{1}{6}\ta1\ta1\ta1.          \nonumber
\eeq
Therefore, the coefficients $\alpha^{x^T}:=\phi(x^T) \in \A$ in (\ref{Zsum}).

Let us briefly dwell on the generalization to decorated
non-planar rooted trees, and to Feynman graphs, following
\cite{overl, dennis}. Every Feynman graph provides a number $r$ of
maximal forests. The integer $r$ counts the number of terms $p_i$,
$i=1,\dots,r$ in the coproduct which are primitive on the rhs, and
in the augmentation ideal on the lhs of the coproduct on graphs. 
If $r>1$, we call the graph
overlapping divergent. It is then mapped to a linear combination
of $r$ decorated rooted trees, where each of those trees has a
root decorated by one of the $p_i$. Iterating this procedure, one
obtains a map from Feynman graphs to decorated rooted trees where
the decorations are provided by subdivergence free skeleton
contributions. Having resolved the overlapping sectors into trees,
one then proceeds as before. 

We close this paper with a study of a simple example on decorated 
rooted trees using two decorations.
The generalization to Feynman graphs including form factor
decompositions for theories with spin is somewhat excessive on the
notational side, but provides no difficulty for the practitioner
of quantum field theory, making full use of the Hopf- and Lie
algebra of Feynman graphs with external structures. See
\cite{BK1, overl, dennis} where examples can be found.

We consider the example of vertices with a decoration $\D$ by 2 
elements $\{a,b\}$. Let us denote them by a vertex $\dta$ and a vertex $\dtb$.
For the Lie bracket (\ref{Liebra}) of these two vertices we get:
\beq
   [Z_{\dta},Z_{\dtb}] = Z_{\dtba} - Z_{\dtab}.     \label{declad1}
\eeq
Note that though we have here the analog of a simple nesting of one graph in
another, this has already a non-vanishing commutator in the Lie
algebra. This fact makes it necessary to include the
BCH-corrections (\ref{BCH1}) already at this level. For the above
example (\ref{declad1}), we have to add the correction
(\ref{1stcorrect}).
 
Let us do the calculation of the counterterm $\phi_{-}$
explicitly for the decorated rooted ladder tree $\dtba$, using the
normal coordinates in (\ref{Zsum}). We have to use $\chi = Z +
\chi^{(1)}$ (\ref{chi}), with $\chi^{(1)}$ given in (\ref{1stcorrect}). 
The infinitesimal character $Z$ generating the character $\phi$ is given 
to order $2$ in terms of the normal coordinates $x^T$ as:
\beq
    Z = \phi(x^{\dta})Z_{\dta} + \phi(x^{\dtb})Z_{\dtb}
                + \phi(x^{\dtab})Z_{\dtab}  + \phi(x^{\dtba})Z_{\dtba},
\eeq
where:
\beqn
    \phi(x^{\dta}) &=& \phi(\dta),\;\;\; \phi(x^{\dtb}) = \phi(\dtb) \nonumber\\
    \phi(x^{\dtab})&=& \phi(\dtab) - \frac{1}{2}\phi(\dta)\phi(\dtb),\;\;\;
    \phi(x^{\dtba})=\phi(\dtba) - \frac{1}{2}\phi(\dta)\phi(\dtb). \label{r3}
\eeqn
And therefore we have for the infinitesimal character $\chi(Z)$ to order $k=1$, i.e including 
the first correction (\ref{1stcorrect}):
\beqn
    \chi &=&  \phi(x^{\dta})Z_{\dta} + \phi(x^{\dtb})Z_{\dtb} + \phi(x^{\dtab})Z_{\dtab}  + \phi(x^{\dtba})Z_{\dtba}\nonumber\\
         & & \hspace{2cm} - \frac{1}{2}\big( [\RB(\phi(x^{\dta})Z_{\dta}),\phi(x^{\dtb})Z_{\dtb}]
                                             + [\RB(\phi(x^{\dtb})Z_{\dtb}),\phi(x^{\dta})Z_{\dta}]\big). \label{chitree}
\eeqn
So that when the counterterm character $\phi_{-}=\EXP(-\RB(\chi))$ is applied to $\dtba$ we get:
\beqn
     \phi_{-}(\dtba) &=& \EXP(-\RB(\chi))(\dtba) \nonumber\\
                     &=& -\RB(\chi)(\dtba) + \frac{1}{2} \RB(\chi) \star \RB(\chi)(\dtba)                          \label{lineA}\\
                     &=& -R\bigg( \phi(\dtba)Z_{\dtba}(\dtba)
                                    -\frac{1}{2} \phi(\dta)\phi(\dtb)Z_{\dtba}(\dtba)          \nonumber\\
                     & & \hspace{1cm}    - \frac{1}{2}\big\{ R(\phi(\dta)Z_{\dta}(\dta))\: \phi(\dtb)Z_{\dtb}(\dtb)
                                           - \phi(\dta)Z_{\dta}(\dta)\: R(\phi(\dtb)Z_{\dtb}(\dtb)) \big\}\bigg)
                                                                                                                   \nonumber\\
                     & & \hspace{2cm}+ \frac{1}{2} R(\phi(\dta)Z_{\dta}(\dta))R(\phi(\dtb)Z_{\dtb}(\dtb))  \label{lineB}\\
                     &=& -R\big( \phi(\dtba) + R(\phi(\dta))\:\phi(\dtb) \big ),                       \label{counter}
\eeqn
which is the correct result.
In line (\ref{lineA}) no higher order terms can appear. In the next line we used relations (\ref{r3}).
From (\ref{lineB}) to the last equality (\ref{counter}) we used the RB relation:
\beqn
   \frac{1}{2} R\big(\phi(\dta)Z_{\dta}(\dta)\big)\hspace{-1cm}& & R\big(\phi(\dtb)Z_{\dtb}(\dtb)\big) +
   \frac{1}{2} R\big( \phi(\dta)Z_{\dta}(\dta)\:\phi(\dtb)Z_{\dtb}(\dtb)\big)                             \nonumber\\
  &=&  \frac{1}{2} R\big\{R\big(\phi(\dta)Z_{\dta}(\dta)\big)\: \phi(\dtb)Z_{\dtb}(\dtb)
                                           +\phi(\dta)Z_{\dta}(\dta)\: R\big(\phi(\dtb)Z_{\dtb}(\dtb)\big)\big\}.
                                                                                                                  \nonumber
\eeqn
\begin{rmk} 
1) Note that now $\chi(u;Z)(T)$, equation (\ref{chiu}), will be a polynomial of degree at
most $m-1$ when acting on decorated trees with $m$ vertices, as
the tree studied above with two vertices differently decorated is
already non-cocommutative under the coproduct.\\[0.2cm]
2) Using the standard example of the QFT $\Phi^3_{6dim}$, the result (\ref{counter}) should be compared 
to the counterterm for the Feynman graph $\graph1$, which has an additional factor two reflecting the fact that 
it is overlapping divergent, $r=2$, and it resolves into two identical rooted trees \cite{overl}.   

\end{rmk}

%
%
\section{Conclusion and Outlook}

In this work we generalized the results of (I) to arbitrary rooted
trees, i.e. we showed how to derive the Birkhoff factorization for
characters of the Hopf algebra of rooted trees. Using the
Rota-Baxter structure underlying the target space of the
characters of a renormalization Hopf algebra, the notion of a
classical r-matrix was introduced on the corresponding Lie algebra
defined on rooted trees. A couple of simple results for
Rota-Baxter algebras were collected which allowed for a
straightforward derivation of the twisted antipode formula,
defined in \cite{Kreim1,CK1} concerning the study of the Hopf
algebraic approach to perturbative QFT. This gives a firm
algebraic basis to any renormalization scheme using an algebraic
Birkhoff decomposition together with a suitable double
construction.

We regard this work as a further step torwards a more interesting
connection to the realm of integrable systems. Sakakibara's result
\cite{Jap} also points into this direction. This connection was
already apparent in \cite{CK3}, in which effectively the grading
operator $Y$ served as a Hamiltonian providing the "scaling
evolution" of the coupling constant, and hence the renormalization
group flow initiated by scaling transformations, and can and
should be worked out for the corresponding flow of many other
physical parameters of interest.
%
%
%

\section*{Acknowledgements}
The first author would like to thank the Ev. Studienwerk for financial support. Also the I.H.\'{E}.S.
and its warm hospitality is greatly acknowledged. We would like to thank Prof. Ivan Todorov,
Prof. Olivier Babelon, and Igor Mencattini for valuable discussions, and helpful comments.
%

%

\end{document}